**Framework for the Treatment And Reporting of Missing data in Observational Studies: The TARMOS framework**

Lee KJ[1,2], Tilling K[3], Cornish RP[3], Little RJA[4], Bell ML[5], Goetghebeur E[6], Hogan JW[7], and Carpenter JR[8] on behalf of the STRATOS initiative.

[1] Murdoch Children's Research Institute, Melbourne, Australia

[2] University of Melbourne, Melbourne, Australia

[3] University of Bristol, Bristol, UK

[4] University of Michigan, US

[5] University of Arizona, US

[6] Ghent University, Belgium

[7] Brown University, US

[8] London School of Hygiene and Tropical Medicine, London, UK

**Abstract**

Missing data are ubiquitous in medical research. Although there is increasing guidance on how to handle missing data, practice is changing slowly and misapprehensions abound, particularly in observational research. We present a practical framework for handling and reporting the analysis of incomplete data in observational studies, which we illustrate using a case study from the Avon Longitudinal Study of Parents and Children. The framework consists of three steps: 1) Develop an analysis plan specifying the analysis model and how missing data are going to be addressed. An important consideration is whether a complete records analysis is likely to be valid, whether multiple imputation or an alternative approach is likely to offer benefits, and whether a sensitivity analysis regarding the missingness mechanism is required. 2) Explore the data, checking the methods outlined in the analysis plan are appropriate, and conduct the pre-planned analysis. 3) Report the results, including a description of the missing data, details on how the missing data were addressed, and the results from all analyses, interpreted in light of the missing data and the clinical relevance. This framework seeks to support researchers in thinking systematically about missing data, and transparently reporting the potential effect on the study results.

**Keywords:** incomplete data, missing data, bias, efficiency, multiple imputation, observational studies, reporting, ALSPAC, STRATOS initiative

## BACKGROUND

Despite recent reviews emphasising the need to minimise missing data during the design stage,[1] missing data remain ubiquitous in medical research. For example, in the Avon Longitudinal Study of Parents And Children (ALSPAC), a transgenerational prospective observational study of 14,500 families in the UK, only 48.2% of children completed the 12 measures collected during adolescence. Electronic routinely-collected clinical datasets, which are increasingly exploited in observational research, are particularly susceptible to missing data because data are collected for clinical reasons, rather than designed research.

Despite an increasing number of papers providing guidance on how to handle missing data,[2-4] practice is changing slowly and misapprehensions abound. This is particularly pertinent in observational research,[5,6] where there is no regulatory framework guiding the analysis and analyses are often adjusted for confounders which can have missing values. Restricting analysis to records with complete data for the analysis model (termed complete case or complete records analysis) is still the most common approach,[7,8] although it is known to result in a loss of power, and in many situations will cause bias. Yet researchers often do not consider the potential impact of missing data on their scientific conclusions.[9] This is despite journals requiring justification for the method used to handle missing data,[10] and tools for assessing the quality of studies having domains referring to how missing data were addressed.

Multiple imputation (MI) is a practical, flexible approach for handling missing data[11] that is becoming increasingly popular.[12,13] Under this approach, missing values are imputed from the predictive distribution of the missing given observed data multiple times. Next, the analysis model is fitted to each `complete' dataset and the results combined using Rubin's rules.[11] A key benefit of MI is that it can readily incorporate *auxiliary* variables (variables predictive of missing values but not in the substantive model) into the imputation step; this can reduce bias and improve efficiency. MI is now available in all the leading statistical software packages. However, this ease of use may result in MI being applied without proper consideration of its appropriateness, and to fundamental mistakes being made.[14,15]

In this paper we propose our Treatment And Reporting of Missing data in Observational Studies (TARMOS) framework, a practical framework for researchers faced with analysing incomplete observational data. We focus on MI because of its flexibility and prominence in the literature, although — as we discuss later — similar principles apply to any approach for handling missing data.

First, we describe a case study from ALSPAC. We then present our framework, illustrating each step in turn. Although we focus on a simple exposure-outcome relationship, the principles underpinning our framework apply quite generally.

## CASE STUDY: THE AVON LONGITUDINAL STUDY OF PARENTS AND CHILDREN

ALSPAC recruited pregnant women living in and around Bristol, England, in the early 1990s. The study has been described previously.[16,17] Briefly, 14,541 women were initially recruited, resulting in 14,062 live births and 13,988 children alive at one year; additional children were enrolled subsequently. ALSPAC has a fully searchable data dictionary and variable search tool (http://www.bristol.ac.uk/alspac/researchers/our-data/). Ethical approval was obtained from the ALSPAC Ethics and Law Committee and the Local Research Ethics Committees.

ALSPAC suffers from attrition and sporadic missingness. Attrition was highest in infancy and late adolescence, and previous analyses have shown that those who continue to participate are more likely to be female and white and less likely to live in low-income households.[16] Our case study explores whether there is a causal relationship between smoking at 14 years and educational attainment at 16 years. This is a modified version of the research question published previously.[18] The analysis used data from 14,684 adolescents - the full cohort less those who died or withdrew consent before 14 years, but there are missing data in all variables required for analysis (except sex). Stata code for the case study is given in Section A of the Supplementary Material.

***Outcome***

Educational attainment score obtained via linkage to the National Pupil Database (https://www.gov.uk/government/collections/national-pupil-database). The score is the percentage of the maximum observed in the data (540 points).

***Exposure***

Participants were asked about smoking via (1) a computerised questionnaire during a clinic assessment (mean age 13.8 years), and (2) a postal questionnaire (mean age 14.1 years). Both included questions about past and current smoking which were used to classify individuals as current smokers or non-smokers.

***Additional variables***

Data were collected on several potential confounding and auxiliary variables capturing education and related social factors at recruitment and previous waves of data collection (Supplementary Table 1).

**THE FRAMEWORK**

Figure 1 outlines our framework. Below we describe the steps of this framework.

< Figure 1 >

***Step 1: Plan the analysis***

When designing a research study, it is important to pre-specify an analysis plan stating the primary and any secondary analyses (prospectively for prospectively collected data). In much observational research, (e.g. our case study), the data will have already been collected. In this context there may be knowledge about the data, including levels of missingness and potential missingness mechanisms, which can be used to develop the analysis plan.

***Step 1a. Identify the substantive research question(s) and plan the statistical analysis***

The first step is to identify the substantive research question(s), i.e. the exposure(s), outcome(s), causal structure, confounders and corresponding analysis model(s). This should (generally) be done without consideration of the missing data. In ALSPAC, the target quantity is the mean difference in educational attainment in smokers versus non-smokers, and our analysis model is a linear regression of educational attainment at 16 years on smoking at 14 years adjusted for confounders outlined in Supplementary Table 1. For simplicity we assume this is a valid analysis model for our question.

***Step 1b. Specify how the missing data will be addressed***

Decisions concerning missing values should be informed by their most plausible contextual cause. For a single incomplete variable, this is often linked to Rubin's typology[19]:

- missing completely at random (MCAR) – missingness does not depend on anything related to the substantive research question, e.g. missingness dependent on wave of data collection in a cross-sectional analysis;
- missing at random (MAR) - the reason for data missingness may depend on its value, but this dependence is broken within strata of (i.e. conditional on) fully observed variables, e.g. missingness on smoking is dependent on smoking status, but not after stratifying by social class (which has no missing data); and
- missing not at random (MNAR) - even within strata of observed variables, missingness still depends on the value itself, e.g. within social strata, missing smoking data depends on smoking status.

Although this classification is useful when there is a single incomplete variable, it is not straight-forward when there are multiple incomplete variables. A more natural way to understand the assumptions and likely impact of missing data in specific analyses where there are multiple imcomplete variables is to use causal diagrams[20-22] (Supplementary Figure 1).

Figure 2 provides an overview of the decision-making process regarding missing data when estimating an exposure-outcome association as in our case study. We propose three key questions to guide the process:

< Figure 2 >

**Q1: Is a complete records analysis likely to give valid inference for the exposure effect?**

This will depend on:

- ***How much information is expected to be lost because of missing values:*** This will depend on which variables are incomplete, the proportion of missing data and the information retained by auxiliary variables. If there is unlikely to be much missing information in the exposure, outcome and key confounders (e.g. if <5% of records are expected to have missing values), it will not make much difference how missing data are handled, irrespective of auxiliary variables, and a complete records analysis might be acceptable.[23] If, however, there is more missing information, e.g. more incomplete records, then MI may be more efficient. This may not be true if there is only missingness in the outcome (dependent variable) and there are no auxiliary variables, as noted below.
- ***What are the likely mechanisms behind missing data:*** There are a range of situations under which a complete records analysis is likely to be unbiased for linear and logistic regression models; these have been outlined in the literature.[20,21,24,25] Importantly, a complete records analysis will be unbiased for estimating a correctly specified exposure-outcome relationship if the reasons for missingness in any variable in the analysis model is not related to the outcome (given the other variables in the analysis model), although it may still be inefficient.

The analysis plan may specify that the strategy for dealing with missing data will depend on the extent of, and reasons for, missing data. For example, the plan could be that if <5% of cases have missing data and there is little evidence that the observed variables are associated with any missingness then a complete records analysis will be used. If, however, ≥5% of cases have missing data and there is evidence that data are not MCAR then MI will be used.

Note, if a complete records analysis is not to be the primary analysis, it can still be useful to conduct such an analysis as a sensitivity analysis that makes a different assumption about the missingness.

In ALSPAC, dropout is associated with many of the covariates in the analysis model (i.e. is not MCAR), and in particular educational attainment (the outcome).[16] Given this, complete records analysis is likely to be biased, and hence would be inappropriate.

**Q2: Is MI (or an alternative inferentially equivalent approach) likely to give a) important bias reduction and/or b) increased precision over a complete records analysis?** This will depend on:

- ***The extent of missing information:*** The more missing information, the greater the potential gains from MI.
- ***Whether there are auxiliary variables that may provide information about the missing values:*** If there are auxiliary variables that are correlated with the incomplete variable(s), including these variables in the imputation model will reduce bias and improve precision over the complete records analysis. In many analyses, there will be a large list of possible auxiliary variables. Typically, it will be best to identify a small number to include, focusing on variables that are strongest, nearly independent predictors of missing values. Variables which predict missingness in one or more variable but are unrelated to the missing values, are of less importance.[26] In selecting auxiliary variables, it is important to consider their completeness; their inclusion is only beneficial if they are observed when the variables of interest are missing.

- ***Which variables are likely to contain missing data:*** If most individuals have complete outcome and exposure data but incomplete confounders then MI can increase the information about the exposure effect. In contrast, there is less to gain if the missingness is in the exposure and/or outcome,[27] unless there are strong auxiliary variables. In the absence of auxiliary variables, MI provides no additional information if only the outcome is incomplete, irrespective of whether covariates are incomplete.

As with all statistical models, an improvement in bias and/or precision with MI is contingent on having an appropriately specified imputation model. In particular, the imputation model needs to be compatible with the substantive model - that is includes the same variables in the same form, including any non-linear terms and interactions.[28] See [29] for a formal description of compatibility.

In ALSPAC, 51% have missing data on smoking status at 14 years, and we expect missingness to be associated with the outcome. There are a number of strong auxiliary variables, such as smoking status at previous and later waves, which are observed when the exposure of interest is missing in some observations. Given this, MI has the potential to reduce bias and improve presicion over a complete records analysis.

**Q3: Is a sensitivity analysis required?** Given that any analysis makes specific (and untestable) assumptions about the missingness mechanism, it is important to explore the robustness of the scientific conclusions to the assumptions.[30] For example, we may wish to carry out an analysis allowing for the fact that data may be MNAR. Another form of sensitivity analysis considers the specification of the imputation models, which relies on numerous subjective decisions. This can be important but, for brevity, we restrict our focus to sensitivity analysis regarding the missingness mechanism.

In ALSPAC, we hypothesised that missingness in smoking at 14 would be associated with smoking itself, conditional on the covariates in the analysis model (i.e. MNAR), hence we specify that we will conduct a sensitivity analysis.

**Step 1c. Provide details on how the MI will be conducted (if required)**

If the analysis plan states that MI (or an alternative MAR method) will be used to handle the missing data, it is important to detail exactly how the analysis will be conducted (including justification) in the analysis plan. For MI this should include: the method of imputation, the variables to be included in the imputation model, the form of variables to be imputed, the nature of the relationships between the variables including any non-linear relationships and interactions, the method of imputation (e.g. multivariate normal imputation,[31] fully conditional specification,[32,33] predictive mean matching etc.), the number of imputations, and the software to be used.

See Section B of the supplementary material for example text for our case study.

**Step 1d. Provide details on how the sensitivity analyses will be conducted (if required)**

Sensitivity analyses can rapidly get very complex, hence it is common to focus on one or two contextually important variables, e.g. the outcome and/or exposure of interest (if a non-trivial proportion of missing values) or the confounder(s) with the largest proportion of missing data.

In a sensitivity analysis we need to change the dependency of the missing values on the other variables, typically the outcome, exposure of interest, or the incomplete variable itself. This can sometimes be done quite simply. For example in ALSPAC, individuals with observed data on smoking at 14 years were less likely to report ever having smoked at 10 and 13 years compared to those with missing data (0.8% vs 3.2% at 10 years and 10.0% vs 29.3% at 13 years). Thus, as an initial, relatively crude, sensitivity analysis, we could explore what happens when smoking is always imputed as `1'.[34,35] If this has limited effect, a more subtle approach is not required. However, when this extreme assumption has a strong effect, we may need to explore more plausible mechanisms.

A simple way to allow different relationships in the complete and incomplete records is using a pattern-mixture approach,[36,37] where e.g. we assume that the value of the variable (or log odds, conditional on the other variables in the imputation model) is different in those observed and unobserved by a value, $\delta$, known as the sensitivity parameter. This is illustrated for our case study in Figure 3. This can be achieved within MI by adding $\delta$ to the imputed values (or linear prediction of the imputed values) within each imputed dataset.[38]

<Figure 3>

Sensitivity analyses rely on external information about how the predictions for missing values differ from those we estimate from the observed values. This information can be elicited from content experts[39] or a tipping-point analysis can be conducted, where a range of values are assumed for $\delta$ to determine how large $\delta$ would need to be to change the overall conclusion.[40] See [38,41-43] for more information on these approaches. The details regarding how the sensitivity analysis will be conducted and how the sensitivity parameters will be obtained should be detailed in the analysis plan.

In ALSPAC, we pre-specified that the sensitivity analysis would be conducted using a pattern-mixture approach, where (after discussion with content experts) we add the fixed log-odds of 0.1, 0.25, 0.5, 1 and 10 (the latter to represent an extreme MNAR mechanism) within the logistic regression model used to impute smoking status using the “offset” option within Stata’s *mi impute chained* command.

***Step 2: Conduct the pre-planned analysis***

**Step 2a. Explore the data**

Once the data have been collected, the first step is to explore the data. This should include:

1. A table showing the proportion of missing data for all variables in the analysis model. Ideally this should be by variable and for the analysis as a whole. It can also be useful to explore the patterns of missing data e.g. which variables are missing together.
2. A table of the observed characteristics for the “complete” versus “incomplete” (or all) participants, or by whether variables with substantial missingness are observed.
3. An assessment of the predictors of missingness, e.g. using a logistic regression model fitted to an indicator for being a complete record, and predictors of missing values i.e. associations with the incomplete variables.

This exploration should be used to judge the methods outlined in the analysis plan, and whether the specified auxiliary variables are likely to be useful.

In ALSPAC, 3,313 of the 14,684 eligible participants (23%) had complete data on all variables required for analysis (Supplementary Table 2). Those with complete records were more likely to be first born, female, have higher educated parents, and have parents who were non-smokers than those with incomplete data (Supplementary Table 3). After adjusting for covariates, educational attainment (the outcome) and smoking at 13 years were associated with being a complete case. This suggests that 1) a complete records analysis would have a much reduced sample size, and 2) the outcome is associated with any missingness. This confirms a complete records analysis will be biased and inefficient and, because we have potentially strong auxiliary variables, MI is likely to reduce bias. It also suggests that the data may be MNAR hence a sensitivity analysis will be important.

**Step 2b. Conduct the analysis as per the analysis plan**

Once satisfied the assumptions made in the analysis plan are acceptable, the next step is to conduct the pre-planned primary, secondary and sensitivity analyses. If the analysis plan needs to be revised in light of the exploratory data analysis, any changes should be acknowledged and justified.

In ALSPAC, data exploration confirmed the methods outlined in the analysis plan are appropriate, hence we proceed with the pre-planned MI and sensitivity analysis.

***Step 3: Reporting***

The methods section of a paper should state how the missing data were addressed in the primary, secondary and sensitivity analyses, including whether this was pre-specified, and any changes made to the pre-specified plan. For each analysis, state the assumptions made (e.g. data are MAR), and provide enough detail for the analysis to be reproducible (outlined in 1c for MI). For the sensitivity analysis, specify how this was conducted (outlined in 1d). Some of these details may appear in the supplementary material.

In the results section, the extent of missing data should be described using the summaries outlined in Step 2a, along with a summary of the reasons for the missing values if possible. Again, some of this information can be included in the supplementary material.

The inference from the various analyses should then be reported and interpreted in light of the missing data and the clinical relevance. Although the main results from secondary and sensitivity analyses should be given in the paper, the full details may be presented in the supplementary material for brevity. If the results from all analyses are similar, the researcher can be reasonably confident that the missing data is having little impact on the inference. In contrast, if there are contextually substantive differences, it is important to suggest an explanation for these, bearing in mind that under the MAR assumption MI should correct at least some of the biases that may arise in a complete records analyses. In this context, it should be made clear which result is likely to be the most accurate based on clinical knowledge, but acknowledge the discrepancy reveals more uncertainty.

Table 1 shows the results from the various analyses of our case study. These results all show strong evidence of a causal relationship between teenage smoking and lower educational attainment at 16 years (assuming the absence of unmeasured confounders and no reverse causality), even in the extreme sensitivity analysis, when we set the sensitivity parameter to 10. Given the similarity of these results we can be reasonably confident this is the true relationship. See Section B in the supplementary material for example text for our case study.

< Table 1 >

## DISCUSSION

We have proposed and illustrated a framework for the planning, analysis and reporting of data from observational studies with incomplete data. The framework places a strong emphasis on pre-specifying the analysis, including how missing data will be handled subject to a priori assumptions regarding the missingness. The full analysis plan could be published or registered for transparency. We highlight the need to assess the validity of the pre-planned methods once the data are available. Finally, we encourage researchers to report the details of the analysis methodology to enable reproducibility, ideally including the statistical code, and to interpret the results based on the clinical relevance and the suspected missingness mechanism.

The proposed framework encourages researchers to exploit information from auxiliary variables to recover information from incomplete observations. However, this relies on the researchers having some insight into the missingness mechanism. Therefore, when designing a study, it is important to identify plausible missingness mechanisms and plan to (i) reduce the extent of missing data during implementation as much as possible (1), and (ii) collect data on potential auxiliary variables.

We have focussed on MI to conduct MAR and MNAR analyses. One attractive feature of MI is that it separates the handling of missing data from the analysis model, so that the decision regarding the analysis model can be made without considering how the missing data will be handled. There are more elaborate ways of conducting MI e.g. using doubly-robust[44] and machine learning methods[45,46] that are not considered here. However, MI is not always the most efficient approach, and can give poor results if not carried out appropriately (i.e. when data are MNAR or using an inappropriate imputation model).[20]

There are a range of alternative methods available for conducting analyses under MAR (or MNAR), such as direct likelihood[47] and full Bayesian analysis.[48] Weighting based methods are another alternative, but present their own challenges.[47,49] MI has the practical advantage of ease of (i) including auxiliary information, (ii) conducting sensitivity analyses, and (iii) handling large datasets. Irrespective of the statistical method chosen, researchers should use the steps presented here, including providing a justification for the analytical approach(es), and enough information to enable readers to repeat the analysis.[50]

In some scenarios, it may be acceptable to only report results from a complete records analysis, e.g. if there is strong justification for data being MCAR or covariates are the only incomplete variables, but this would need careful justification.[24]

We have focussed on the simple scenario of estimating an exposure-outcome relationship adjusted for confounders. The same principles would, however, apply for more complex analyses e.g. using propensity scores.[51] However, if the analysis involves particularly complex analysis models, e.g. splines or hierarchical models, or specific forms of missing data, e.g. in linkage data, then conducting an MAR analysis may require more sophisticated methods than presented here.[29]

Finally, we propose using simple sensitivity analyses if required. Firstly, this can be difficult to judge as it is not possible to determine from the observed data whether data are MAR or MNAR. And secondly, sensitivity analyses can become complex if there is missingness in multiple variables as it is not always clear what assumptions to make under MNAR in such analyses. Methods have been developed to conduct complex sensitivity analyses, for example not at random fully conditional specification (NARFCS),[52,53] and for elicitation of sensitivity parameters,[54] although these are beyond the scope of this manuscript.

In summary we have proposed an accessible framework for planning, analysis and reporting studies with missing data. By following the framework researchers will be encouraged to think carefully about missing data and the assumptions made during analysis, and be more transparent about the potential effect on the study results. If adopted, this framework will improve the standards and increase confidence in the reliability and reproducibility of published results.[55]

**Acknowledgements**

This work was supported by the Australian National Health and Medical Research Council (career development fellowship 1127984 to KJL). The UK Medical Research Council and Wellcome (Grant ref: 102215/2/13/2) and the University of Bristol provide core support for ALSPAC. KT and RC work in the Integrative Epidemiology Unit which receives funding from the UK Medical Research Council and the University of Bristol (MC_UU_00011/3). JC is supported by UK Medical Research Council, grants MC UU 12023/21 and MC UU 12023/29. This publication is the work of the authors and Katherine Lee and Rosie Cornish will serve as guarantors for the contents of this paper.

We are extremely grateful to all the families who took part in this study, the midwives for their help in recruiting them, and the whole ALSPAC team, which includes interviewers, computer and laboratory technicians, clerical workers, research scientists, volunteers, managers, receptionists and nurses. We would also like to thank the STRengthening Analytical Thinking for Observational Studies (STRATOS) publication panel for providing feedback on this manuscript. The international STRATOS initiative (http://stratos-

initiative.org) aims to provide accessible and accurate guidance documents for relevant topics in the design and analysis of observational studies.

We would like to thank Dan Smith and Margarita Moreno-Betancur for their helpful comments on the manuscript.

Conflict of interest: none declared.

**References**

1. Little RJA, Cohen ML, Dickersin K, et al. The design and conduct of clinical trials to limit missing data. *Statistics in Medicine.* 2012;31(28):3433–3443

2. Little RJA, D'Agostino R, Cohen ML, et al. The Prevention and Treatment of Missing Data in Clinical Trials. *New England Journal of Medicine.* 2012;367:1355-1360.

3. National Research Council. *The prevention and treatment of missing data in clinical trials (https://www.nap.edu/read/12955/chapter/1).* Washington, DC2010.

4. Hogan JW, Roy J, Krokontzelou C. Tutorial in biostatistics: handling drop-out in longitudinal studies. *Statistics in Medicine.* 2004;23:1455–1497.

5. Klebanoff MA, Cole SR. Use of Multiple Imputation in the Epidemiologic Literature. *American Journal of Epidemiology.* 2008;168(4):355-357.

6. Kalaycioglu O, Copas A, King M, Omar RZ. A comparison of multiple imputation methods for handling missing data in repeated measurements observational studies. *Journal of the royal Statistical Society Series A.* 2016;179:683–706.

7. Bell ML, Fiero M, Horton NJ, Hsu C. Handling missing data in RCTs; a review of the top medical journals. *BMC Medical Research Methodology.* 2014;14:118.

8. Eekhout I, de Boer RM, Twisk JW, de Vet HC, Heymans MW. Missing data: a systematic review of how they are reported and handled. *Epidemiology.* 2012;23(5):729-732.

9. Karahalios A, Baglietto L, Carlin JB, English DR, Simpson JA. A review of the reporting and handling of missing data in cohort studies with repeated assessment of exposure measures. *BMC Medical Research Methodology.* 2012;12:96.

10. Ware JH, Harrington D, Hunter DJ, D'Agostino RB. Missing data. *New England Journal of Medicine.* 2012;367(14):1353-1354.

11. Rubin DB. *Multiple imputation for nonresponse in surveys.* New York: Wiley; 1987.

12. Mackinnon A. The use and reporting of multiple imputation in medical research - a review. *J Intern Med.* 2010;268(6):586-593.

13. Rezvan PH, Lee KJ, Simpson JA. The rise of multiple imputation: A review of the reporting and implementation of the method in medical research. *BMC Medical Research Methodology.* 2015;15:30.

14. Hippisley-Cox J, Coupland C, Vinogradova Y, Robson J, May M, Brindle P. QRISK—authors' response [electronic response]. *BMJ.* 2007.

15. Hippisley-Cox J, Coupland C, Vinogradova Y, Robson J, May M, Brindle PB. Derivation and validation of QRISK, a new cardiovascular disease risk score for the United Kingdom: prospective open cohort study. *BMJ.* 2007;335(7611):136.

16. Boyd A, Golding J, Macleod J, et al. Cohort Profile: The 'Children of the 90s'—the index offspring of the Avon Longitudinal Study of Parents and Children. *International Journal of Epidemiology.* 2013;42(1):111–127. .

17. Fraser A, Macdonald-Wallis C, Tilling K, et al. Cohort Profile: The Avon Longitudinal Study of Parents and Children: ALSPAC mothers cohort. *International Journal of Epidemiology.* 2013;42:97-110.

18. Stiby AI, Hickman M, Munafò MR, Heron J, Yip VL, Macleod J. Adolescent cannabis and tobacco use and educational outcomes at age 16: birth cohort study. *Addiction.* 2015;110(4):658–668.

19. Rubin DB. Inference and missing data. *Biometrika* 1976;63:581-592.

20. Hughes RA, Heron J, Sterne JAC, Tilling K. Accounting for missing data in statistical analyses: multiple imputation is not always the answer *International Journal of Epidemiology.* 2019:dyz032.

21. Moreno-Betancur M, Lee KJ, Leacy FP, White IR, Simpson JA, Carlin JB. Canonical Causal Diagrams to Guide the Treatment of Missing Data in Epidemiologic Studies *American Journal of Epidemiology.*187(12):2705–2715.

22. Daniel RM, Kenward MG, Cousens SN, De Stavola BL. Using causal diagrams to guide analysis in missing data problems. *Stat Meth Med Res.* 2012;21(3):243-256.

23. Madley-Dowd P, Hughes R, Tilling K, Heron J. The proportion of missing data should not be used to guide decisions on multiple imputation. *J Clin Epidemiol.* 2019:S0895-4356(0818)30871-30870. [Epub ahead of print].

24. White IR, Carlin JB. Bias and efficiency of multiple imputation compared with complete-case analysis for missing covariate values. *Statistics in Medicine.* 2010;29(28):2920-2931.

25. Bartlett JW, Harel O, Carpenter JR. Asymptotically Unbiased Estimation of Exposure Odds Ratios in Complete Records Logistic Regression. *Am J Epidemiol.* 2015;182(8):730-736.

26. Spratt M, Carpenter J, Sterne JA, Carlin JB, Heron, J., Henderson J, Tilling K. Strategies for multiple imputation in longitudinal studies. *American Journal of Epidemiology.* 2010;172(4):478-487.

27. Lee KJ, Carlin JB. Recovery of information from multiple imputation: a simulation study. *Emerging themes in epidemiology.* 2012;9(1):3.

28. Tilling K, Williamson EJ, Spratta M, Sterne JAC, Carpenter JR. Appropriate inclusion of interactions was needed to avoid bias in multiple imputation. *Journal of Clinical Epidemiology.* 2016;80:107–115.

29. Bartlett JW, Seaman SR, White IR, Carpenter JR, for the Alzheimer's Disease Neuroimaging Initiative. Multiple imputation of covariates by fully conditional specification: accomodating the substantive model. *Statistical Methods in Medical Research.* 2015;24(4):462-487.

30. Hogan JW, Daniels MJ, Hu L. A Bayesian Perspective on Assessing Sensitivity to Assumptions about Unobserved Data. In: Molenberghs G, Fitzmaurice G, Kenward MG, Tsiatis A, Verbeke G, eds. *Handbook of missing data methodology.* Chapman and Hall/CRC Press; 2014.

31. Schafer JL. *Analysis of Incomplete Multivariate Data.* London: Chapman & Hall; 1997.

32. Raghunathan TE, Lepkowski JM, Van Hoewyk J, Solenberger P. A multivariate technique for multiply imputing missing values using a sequence of regression models. *Survey Methodology.* 2001;27:85-95.

33. Azur MJ, Stuart EA, Frangakis C, Leaf PJ. Multiple imputation by chained equations: what is it and how does it work? *Int J Methods Psychiatr Res.* 2011;20(1):40-49.

34. Carpenter J. Multiple Imputation-Based Sensitivity Analysis. In: Wiley, ed. *StatsRef: Statistics Reference Online, pp.1-18.*2019:1-18.

35. Hedeker D, Mermelstein RJ, Demirtas H. Analysis of binary outcomes with missing data: missing= smoking, last observation carried forward, and a little multiple imputation. *Addiction.* 2007;102(10):1564-1573.

36. Little RJA. Pattern-mixture models for multivariate incomplete data. *Journal of the American Statisical Association.* 1993;88:125-134.

37. Diggle P, Kenward MG. Informative drop-out in longitudinal data analysis. *Journal of the Royal Statistical Society: Series C (Applied Statistics).* 1994;43(1):49-93.

38. Yuan Y. *Sensitivity Analysis in Multiple Imputation for Missing Data.* Vol Paper SAS270-2014: SAS Intitute Inc; 2014.

39. White IR. Chapter 21. Sensitivity analysis: the elicitation and use of expert opinion. In: Molenberghs G, Fitzmaurice G, Kenward MG, Tsiatis A, Verbeke G, eds. *Handbook of Missing Data Methodology.* London: Chapman and Hall/CRC; 2015.

40. Molenberghs G, Fitzmaurice G, Kenward MG, Tsiatis A, Verbeke G. *Handbook of Missing Data Methodology.* Chapman and Hall/CRC; 2014.

41. Little RJA. A class of pattern-mixture models for normal incomplete data. *Biometrika* 1994;81(3):471-483.

42. VanBuuren S, Boshuizen HC, Knook DL. Multiple imputation of missing blood pressure covariates in survival analysis. *Statistics in Medicine.* 1999;18:681-694.

43. Van Buuren S. *Flexible Imputation of Missing Data.* Hoboken: CRC Press; 2012.

44. Carpenter JR, Kenward MG, Vansteelandt S. A comparison of multiple imputation and inverse probability weighting for analyses with missing data. *Journal of the Royal Statistical Society, Series A (Statistics in Society).* 2006;169:571-584.

45. Loh W, Eltinge J, Cho M, Li Y. Classification and regression tree methods for incomplete data from sample surveys. *Statistica Sinica.* 2019;29:431-453.

46. Shah AD, Bartlett J, Carpenter J, Nicholas O, Hemingway H. Comparison of random forest and parametric imputation models for imputing missing data using MICE: a CALIBER study. *American journal of epidemiology.* 2014;179:764-774.

47. Little RJA, Rubin DB. *Statistical Analysis with Missing Data.* Chichester: Wiley; 1987.

48. Daniels MJ, Hogan JW. *Missing Data in Longitudinal Studies: Strategies for Bayesian Modeling and Sensitivity Analysis.* Chapman & Hall/CRC; 2008.

49. Robins JM, Rotnitzky A, Zhao LP. Analysis of semiparametric regression models for repeated outcomes in the presence of missing data. *J Am Statist Ass.* 1995;90:106-129.

50. von Elm E, Altman DG, Egger M, et al. Strengthening the reporting of observational studies in epidemiology (STROBE) statement: guidelines for reporting observational studies in epidemiology. *BMJ.* 2007;335:806-808.

51. Victorian Centre for Biostatistics. http://www.vicbiostat.org.au/.

52. Leacy FP. *Multiple imputation under missing not at random assumptions via fully conditional specification [dissertation]*.

53. Tompsett DM, Leacy F, Moreno-Betancur M, Heron J, White IR. On the use of the not-at-random fully conditional specification (NARFCS) procedure in practice. *Statistics in Medicine.* 2018;37(15):2338–2353.

54. Mason AJ, Gomes M, Grieve R, Ulug P, Powell JT, Carpenter J. Development of a practical approach to expert elicitation for randomised controlled trials with missing health outcomes: Application to the IMPROVE trial. *Clinical Trials.* 2017;14(4):357–367.

55. Sterne JAC, Savović J, Page MJ, et al. RoB 2: a revised tool for assessing risk of bias in randomised trials. *BMJ.* 2019;366:l4898.

Figure 1: The framework

**1. Plan the analysis**

a) What is the analysis model of interest assuming there are no missing data?

b) How are missing data going to be handled?

- Is a complete records analysis likely to be valid?
- Is MI likely to offer benefits over a complete records analysis?
- Is a sensitivity analysis required, and if so how it should be framed?

**2. Conduct the analysis**

a) Explore the data – are the features of the data consistent with the expectation outlined in the analysis plan?

b) Conduct the analysis as per the plan – any amendments to the analysis plan should be acknowledged and justified.

**3. Report the analysis**

a) Describe the missing data

b) Describe how the missing data were handled in the primary and secondary analyses and provide a justification for the selected approach(es), including details of any non-standard issues e.g. non-linear terms, interactions etc.

c) Report the results from all of the analyses and interpret in light of the missing data and the clinical relevance, commenting on any substantial differences in the inference if relevant

Figure 2: Flowchart for selecting an appropriate method to handle the missing data

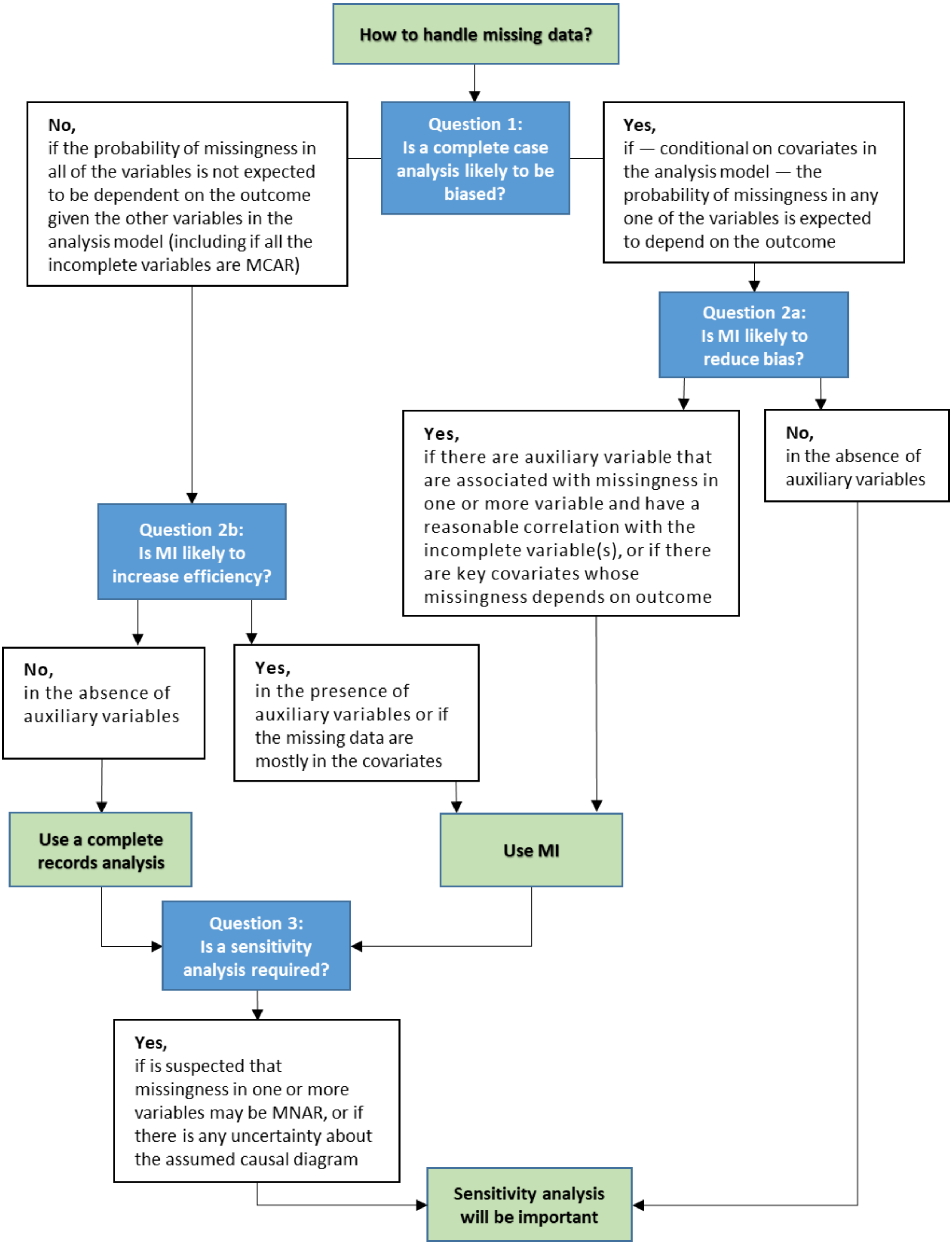

Figure 3: Sensitivity analysis where we vary the imputation distribution of the incomplete smoking exposure: (A) Relationship between the incomplete exposure and the other variables in the dataset in the observed data; (B) Relationship between the incomplete exposure and the other variables in the dataset assumed in the complete and incomplete cases under MAR, namely that the relationship is the same for all cases; (C) Relationship between the incomplete exposure and the other variables in the dataset assumed in the complete and incomplete cases in the (MNAR) sensitivity analysis where we allow these relationships to be different in those with complete and incomplete data.

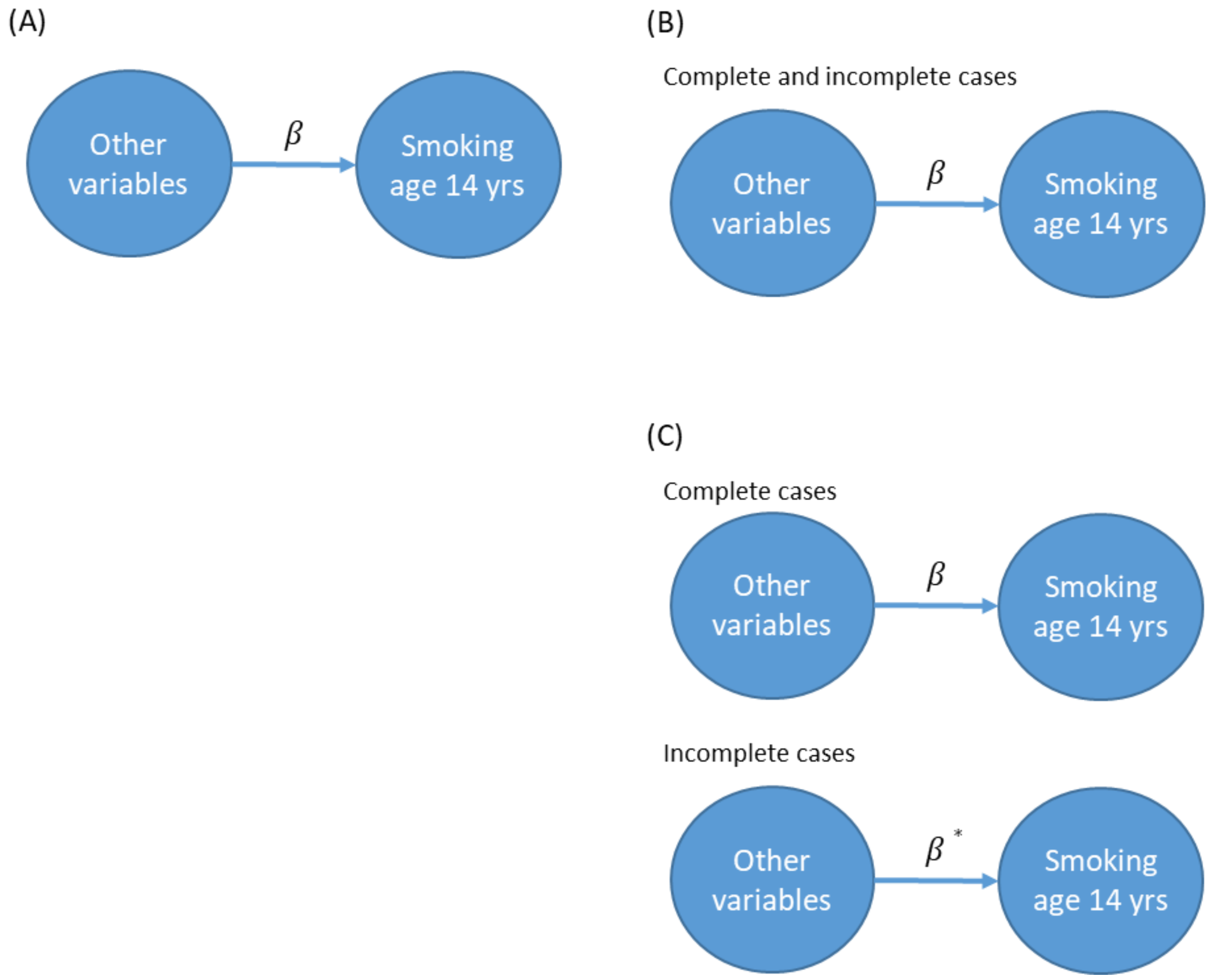

Note, in practice we can never know $\beta^*$. Instead we specify a sensitivity parameter, $\delta$, or a range of $\delta$'s, which represent the hypothesised difference between $\beta$ and $\beta^*$.

**Table 1: Analysis of the relationship between smoking at 14 years and educational attainment at 16 years**

| Method of Analysis | Regression coefficient (95% CI) | p | % of missing smoking values imputed as 'smokers" |
|---|---|---|---|
| Primary analysis: Multiple imputation | -10.8 (-12.2, -9.4) | <0.001 | 13.3 |
| Complete records analysis | -7.9 (-9.1, -6.7) | <0.001 | N/A |
| Sensitivity Analysis – sensitivity parameter = 0.1 | -10.9 (-12.4, -9.4) | <0.001 | 14.2 |
| Sensitivity Analysis – sensitivity parameter = 0.25 | -11.0 (-12.3, -9.6) | <0.001 | 15.5 |
| Sensitivity Analysis – sensitivity parameter = 0.5 | -11.0 (-12.3, -9.6) | <0.001 | 18.1 |
| Sensitivity Analysis – sensitivity parameter = 1 | -10.7 (-11.8, -9.6) | <0.001 | 24.2 |
| Sensitivity Analysis – sensitivity parameter = 10 | -4.3 (-4.7, -3.8) | <0.001 | 99.8 |

**Supplementary material**

## Section A: Stata code to reproduce the analysis for the Avon Longitudinal Study of Parents and Children (ALSPAC) case study

```
clear all
capture log close
version 15
use smoke_attain
egen miss=rowmiss(ks4pct smoke14b sex parity mumed daded dadsmoke ///
      smoke ks2pct sdqtot81)
replace miss=1 if miss>1
gen complete=1-miss
save smoke_attain, replace

* Summarise characteristics of full sample and complete cases
tab1 smoke14b sex parity mumed daded dadsmoke smoke
summ ks2pct ks4pct sdqtot81, detail
tab1 smoke14b sex parity mumed daded dadsmoke smoke if complete==1
summ ks2pct ks4pct sdqtot81 if complete==1, detail

* Predictors of being a complete case
logistic complete sex
logistic complete i.parity
logistic complete i.mumed
logistic complete i.daded
logistic complete i.smoke
logistic complete dadsmoke
logistic complete sdqtot81
logistic complete ks2pct
logistic complete smoke14b
logistic complete ks4pct

*Complete case analysis
regress ks4pct smoke14b sex i.parity i.mumed i.daded dadsmoke ///
      i.smoke ks2pct sdqtot81

*Multiple imputation
mi set mlong
mi register regular sex
mi register imputed ks4pct smoke14b parity mumed daded dadsmoke ///
   smoke ks2pct sdqtot81 car sclasshigh housing bfduration rooms ///
   eversmoke10 eversmoke13 fsmoke15 iq8 behave57
mi impute chained ///
   (regress, omit((ks4pct) (ks2pct)) include((ks4pct^3) (ks2pct^2))) iq8
///
   (regress, omit((iq8)) include((iq8^(1/3)))) ks4pct ///
   (regress, omit((iq8)) include((iq8^(1/2)))) ks2pct ///
   (pmm, knn(5)) rooms sdqtot81 behave57 ///
   (logit) bfduration housing sclasshigh car dadsmoke eversmoke10 ///
   eversmoke13 smoke14b ///
   (mlogit) parity daded mumed smoke fsmoke15 ///
   = sex, burnin(20) add(100) rseed(5432127) augment
save smkattain_impute100, replace
mi estimate: regress ks4pct smoke14b sex i.parity i.mumed i.daded ///
```

```
   dadsmoke i.smoke ks2pct sdqtot81

*Sensitivity analysis
use smoke_attain, clear
foreach delta in 0.1 0.25 0.5 1 10 {
      use smoke_attain
      gen msmk=mi(smoke14b)
      gen sens=`delta'*msmk
      mi set mlong
      mi register regular sex
      mi register imputed ks4pct smoke14b bfduration parity car ///
         sclasshigh housing mumed daded dadsmoke smoke ks2pct eversmoke10
///
         eversmoke13 eversmoke15 fsmoke15 iq8 rooms behave57 sdqtot81
      mi impute chained ///
         (regress, omit((ks4pct) (ks2pct)) include((ks4pct^3) (ks2pct^2)))
         iq8 ///
         (regress, omit((iq8)) include((iq8^(1/3)))) ks4pct ///
         (regress, omit((iq8)) include((iq8^(1/2)))) ks2pct ///
         (pmm, knn(5)) rooms sdqtot81 behave57 ///
         (logit) sclasshigh car dadsmoke eversmoke10 eversmoke13 ///
         bfduration housing ///
         (logit, offset(sens)) smoke14b ///
         (mlogit) parity daded mumed smoke fsmoke15 ///
         = sex, burnin(20) add(10) rseed(5432127) augment
      mi estimate: regress ks4pct smoke14b sex i.parity i.mumed i.daded ///
         dadsmoke i.smoke ks2pct sdqtot81
      save misens`delta'_imp20, replace
      clear
}
```

**Section B: Example of the pre-specified statistical analysis plan, and the methods, results, conclusions for a paper for the ALSPAC case study**

**1) Statistical Analysis Plan**

The causal relationship between teenage smoking (at 14 years) and educational attainment at age 16 years will be assessed using a linear regression of educational attainment at age 16 years on smoking at age 14 years adjusted for the following confounders: child's sex; parity; maternal and paternal smoking status and educational level (O level/Certificate of Secondary Education/vocational, A level[1], and degree or higher); the child's total score on the Strength and Difficulties Questionnaire (SDQ), a measure of behavioural difficulties, measured using a parent-completed questionnaire at age 81 months; and attainment score at age 11 years (ranging from 0 to 280, converted to a percentage).

We expect there to be relatively high rates of missingness, particularly in smoking status, the exposure of interest. Missing data in ALSPAC has previously been shown to be associated with many of the covariates in the analysis model (i.e. is not MCAR). In particular, attrition and non-response have been shown to be associated with educational attainment (the outcome measure of interest in this analysis), with those with lower attainment being less likely to respond {Boyd, 2013 #506}. There are also a number of potentially useful auxiliary variables, such as smoking status at previous and later waves for imputing the smoking exposure and IQ for imputing the outcome, educational attainment. Given this, MI will be the primary method of analysis at it has the potential to reduce bias and improve precision over a complete records analysis. However, we will also conduct a complete records analysis as a comparison.

We also hypothesise that missingness in smoking at age 14 years will be associated with smoking itself, conditional on the covariates in the analysis model (i.e. MNAR), hence we will conduct a sensitivity analysis. The sensitivity analysis will be conducted using a pattern-mixture approach, where we will apply a range of sensitivity parameters within the logistic regression model used to impute smoking status. This will be incorporated using the "offset" option within Stata's *mi impute chained* command, where (after discussion with content experts) we will add the fixed amounts of 0.1, 0.25, 0.5, 1 and 10 (the latter to represent a very extreme MNAR mechanism) to the imputed log odds of smoking to increase the log odds of smoking among those with missing smoking data.

MI will be conducted using fully conditional specification applied to all the variables in the analysis model, as well as the auxiliary variables: smoking (ever smoked) reported by the child at study clinics at age 10 and 13 years; frequency of smoking at 15 years (never, < daily and daily) reported at a study clinic; IQ measured at a study clinic at 8 years using the Wechsler Intelligence Scale for Children 3rd edition (WISC-III; ref); child behaviour score at age 57 months, generated from five questions included on a parent-completed questionnaire (frequency the child bullies other children, is disobedient, tells lies, takes things belonging to others, fights with other children, each measured using a Likert-type scale from never to always); duration of breastfeeding (<3 months, 3+ months); and additional measures of socio-economic position measured during pregnancy: family occupational social class (classified as manual vs. non-manual), number of rooms in home (excluding bathrooms), housing tenure (owned/mortgaged vs rented/other) and car ownership. It is known that the attainment scores (at age 16 years, the outcome of interest, and at age 11 years, covariate) are not linearly associated with IQ. Therefore, fractional polynomials will used to obtain the best fitting non-linear model for these relationships, which will then be incorporated into the imputation model. These variables will be included as linear variables when imputing other variables. Predictive mean matching, selecting from the 5 nearest neighbours, will be used to impute both behaviour scores and number of rooms in the home because these variables are positively skewed and take on only positive values. The remaining covariates will be imputed using either logistic or multinomial logistic regression, as applicable. MI will be conducted using Stata's *mi impute chained command*; 100 datasets will be imputed with a burn-in of 20 iterations.

[1] CSEs (Certificate of Secondary Education) and O levels were qualifications taken at age 16 – now replaced by GCSEs (General Certificate of Secondary Education) in England, Wales and Northern Ireland. A levels are exams taken at age 18 in these countries.

## 2) Methods, results, conclusions for a published paper

### Methods

We first summarised the data for the variables relevant to this analysis, including the amount of missing data, both overall and for each variable in turn.

The causal relationship between teenage smoking (at 14 years) and educational attainment at age 16 years was then assessed using a linear regression of educational attainment at age 16 years on smoking at age 14 years adjusted for the following confounders: child's sex; parity; maternal and paternal smoking status and educational level (certificate of secondary education/vocational, O level, A level, and degree or higher); the child's total score on the Strength and Difficulties Questionnaire (SDQ), a measure of behavioural difficulties, measured using a parent-completed questionnaire at age 81 months; and attainment score at age 11 years (ranging from 0 to 280, converted to a percentage). As per our pre-specified analysis plan, MI was used to address the missing data as the primary analysis, although as a sensitivity analysis we also present the data from a complete records analysis. Finally, we conducted a sensitivity analysis where we assumed an MNAR missingness mechanism for the exposure of interest (smoking at age 14 years). The sensitivity analysis was conducted using a pattern-mixture approach, where we applied a range of sensitivity parameters (0.1, 0.25, 0.5, 1 and 10), the latter to represent a very extreme MNAR mechanism), as an "offset" within the logistic regression model used to impute smoking status.

All MI analyses were conducted using fully conditional specification applied to all the variables in the analysis model, as well as the auxiliary variables (ever smoked at 10 and 13 years, frequency of smoking at 15 years, IQ at 8 years, child behaviour score at age 57 months, duration of breastfeeding, family occupational social class, number of rooms in home, housing tenure, and car ownership). Attainment scores (at age 16 years, the outcome of interest, and at age 11 years, covariate) were imputed using a non-linear relationship with IQ based on fractional polynomials, which resulted in attainment at age 16 years (the outcome) being imputed using linear regression dependent on the cube root of IQ and the age 11 years attainment score being imputed from the square root of IQ. Similarly, IQ was imputed from the attainment score at age 16 years cubed and attainment score at age 11 years squared. These variables were included as linear variables when imputing other variables. Predictive mean matching, selecting from the 5 nearest neighbours, was used to impute both behaviour scores and number of rooms in the home. The remaining covariates were imputed using either logistic or multinomial logistic regression, as applicable. In each case, MI was conducted using Stata's *mi impute chained command*; 100 datasets were imputed with a burn-in of 20 iterations.

### Results

Supplementary Table 2 presents a summary of the variables relevant to this research question. Importantly, only 22% of participants have complete data on all the variables required for the substantive analysis. Participants with complete records were more likely to be first born, female, to have more highly educated parents, and to have parents who were non-smokers than those with incomplete data (Supplementary Tables 2 and 3). Importantly, after adjusting for covariates, educational attainment (the outcome) (odds ratio [OR]=1.37 95% confidence interval [CI] 1.27, 1.47 per 10% increase in attainment), and smoking at 13 years (OR=0.35 95% CI 0.26, 0.47) were associated with being a complete case. This confirms a complete records analysis will be biased and, because we have auxiliary variables likely to be strongly associated with smoking and attainment, this justifies the use of MI as the primary analysis.

Table 1 in the main manuscript presents the estimated mean difference in attainment score comparing those who smoked to those who did not obtained from the primary analysis (MI), the complete records analysis and the sensitivity analyses. All of these results show that smoking at age 14 years is associated with lower educational attainment at age 16 years. This is even the case for the extreme sensitivity analysis, when we set the sensitivity parameter to 10.

**Conclusions**

There is a causal relationship between smoking at age 14 years and lower educational attainment at age 16 years under the assumptions of no residual confounding and no reverse causality.

## Section C: Supplementary tables and figures

**Supplementary Figure 1: Causal diagram for the ALSPAC case study**

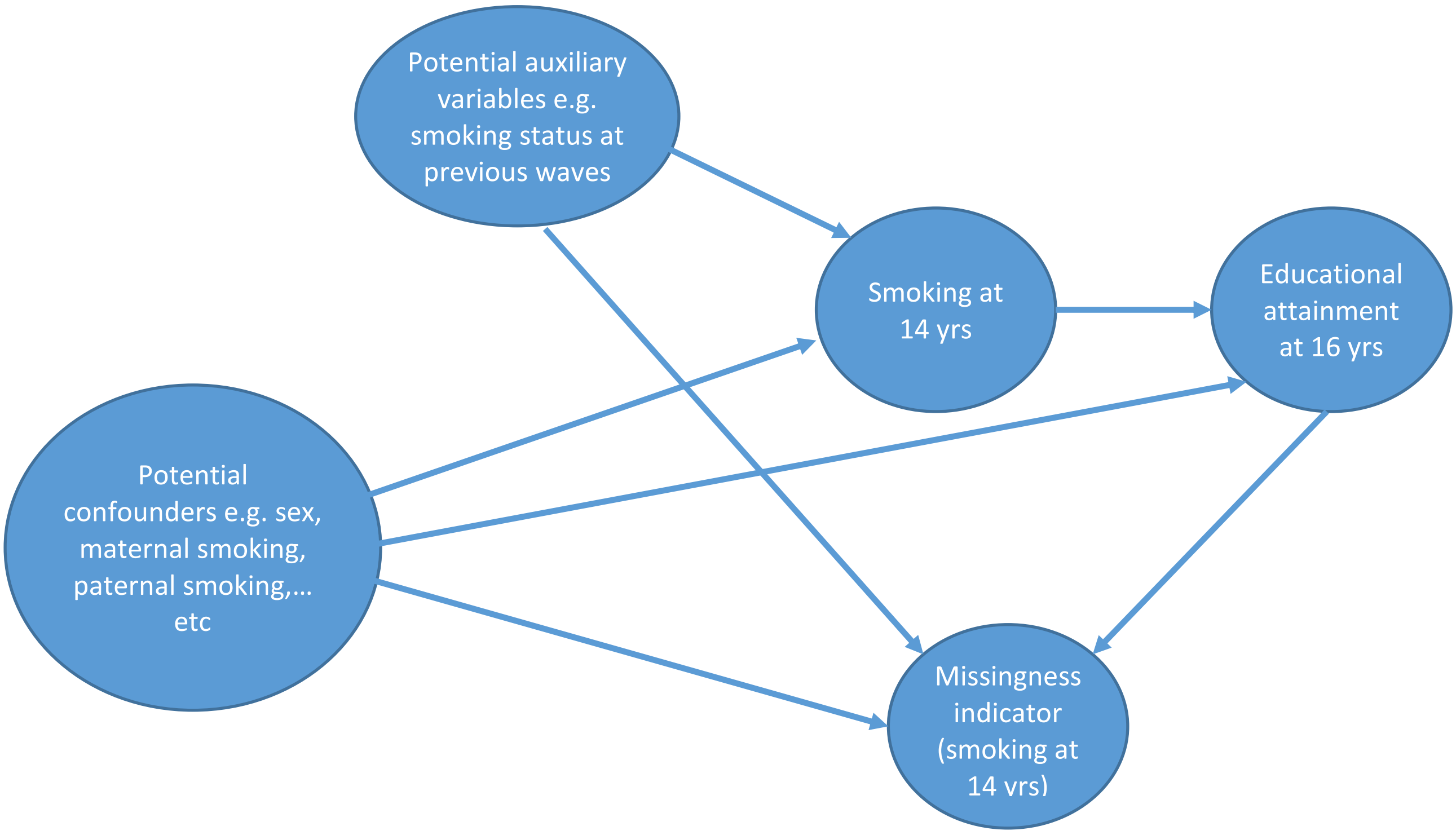

Note, this figure illustrates the fact that we expect missingness to depend on the outcome of interest, educational attainment, and the presence of potential auxiliary variables that are both associated with missingness and with the incomplete exposure variable (smoking age 14 years)

**Supplementary Table 1: Variables of interest in the ALSPAC case study**

| Variable | Variable name | Values |
|---|---|---|
| Educational attainment score at 16 years (outcome) | ks4pct | 0-100% |
| Smoking at 14 years (exposure) | smoke14b | 0=non-smoker<br>1=current smoker |
| **Confounders** | | |
| Child sex | sex | 0=male; 1=female |
| Parity | parity | 0, 1, 2, 3+ |
| Maternal smoking status | smoke | 1 = never<br>2 = yes, but not in current pregnancy<br>3 = yes, including in pregnancy |
| Paternal smoking status | dadsmoke | 0 = never<br>1 = current or previous smoker |
| Maternal educational level | mumed | 0 = O level/CSE/vocational<br>1 = A level<br>2 = degree or higher |
| Paternal educational level | daded | As above |
| Behavioural difficulties score at 81 months | sdqtot81 | 0-40 |
| Attainment score at 11 years | ks2pct | 0-100% |
| **Auxiliary variables** | | |
| Smoking age 10 years | eversmoke10 | 0 = never smoked<br>1 = current or previous smoker |
| Smoking age 13 years | eversmoke13 | 0 = never smoked<br>1 = current or previous smoker |
| Frequency of smoking at 15 years | fsmoke15 | 0 = never<br>1 = < daily<br>2 = daily |
| IQ age 8 years | iq8 | 45-151 (range in data) |
| Behaviour score at 57 months | behave57 | 0-20 |
| Duration of breastfeeding | bfduration | 0 = never/<3 months<br>1 = 3+ months |
| Number of rooms in home (excluding bathrooms) during pregnancy | rooms | 0 to 9 |
| Family occupational social class (higher of maternal and paternal) | sclasshigh | 0 = non-manual<br>1 = manual |
| Car ownership | car | 0 = yes; 1 = no |
| Housing tenure | housing | 0 = mortgaged/owned<br>1 = private rented / other |

**Supplementary Table 2: Summary of the variables in the analysis model for the ALSPAC case study including the amount of data available for each variable, and a summary of the characteristics for the enrolled sample and those with complete records.**

| Characteristic | | Available data (n=14,684) N (%) | Enrolled singletons and twins alive at one year and not withdrawn (n=14,684)[1] | Complete records (n=3,313) |
|---|---|---|---|---|
| Sex | Male<br>Female | 14,684 (100%) | 7,536 (51%)<br>7,148 | 1,559 (47%)<br>1,754 |
| Parity | 0<br>1<br>2+ | 12,924 (88%) | 5,770 (45%)<br>4,539 (35%)<br>2,615 (20%) | 1,628 (49%)<br>1,181 (36%)<br>504 (15%) |
| Mother's education | O level/lower<br>A level<br>Degree/higher | 12,412 (85%) | 8,022 (65%)<br>2,791 (22%)<br>1,599 (13%) | 1,800 (54%)<br>932 (28%)<br>581 (18%) |
| Father's education | O level/lower<br>A level<br>Degree/higher | 10,717 (73%) | 5,445 (51%)<br>3,104 (29%)<br>2,168 (20%) | 1,473 (44%)<br>1,054 (32%)<br>786 (24%) |
| Mother's smoking | Never smoked<br>Smoked, not in pregnancy<br>Smoking in pregnancy | 13,242 (90%) | 6,413 (48%)<br>3,584 (27%)<br>3,245 (25%) | 1,958 (59%)<br>934 (28%)<br>421 (13%) |
| Paternal smoking (ever smoked) | No<br>Yes | 10,690 (73%) | 4,419 (41%)<br>6,271 | 1,624 (49%)<br>1,689 |
| Behavioural difficulties score at 81 months | Median (IQR) | 7,289 (50%) | 6 (4-10) | 6 (4-9) |
| Attainment score at 11 years | Mean (SD) | 11,813 (80%) | 65% (16%) | 71% (14%) |
| Smoking at 14 years | No<br>Yes | 7,211 (49%) | 6,762 (94%)<br>449 (6%) | 3,123 (94%)<br>190 (6%) |
| Outcome: attainment score | Mean (SD) | 12,020 (82%) | 58% (18%) | 67% (13%) |

Note, there are 3,153 participants who have complete data on all of these variables required for analysis (21% of the original 14,684).

[1] Denominators vary because the variables come from different sources/questionnaires and have different completion rates.

**Supplementary Table 3: Predictors of being a complete case in the ALSPAC case study (n=14,684)[1]**

| Characteristic | | Crude odds ratio (95% confidence interval) | Area under the curve |
|---|---|---|---|
| Sex | Male<br>Female | 1.00<br>1.25 (1.15, 1.35) | 0.53 |
| Parity | 0<br>1<br>2+ | 1.00<br>0.89 (0.81, 0.98)<br>0.61 (0.54, 0.68) | 0.54 |
| Mother's education | O level/lower<br>A level<br>Degree/higher | 1.00<br>1.73 (1.58, 1.90)<br>1.97 (1.76, 2.21) | 0.57 |
| Father's education | O level/lower<br>A level<br>Degree/higher | 1.00<br>1.39 (1.26, 1.53)<br>1.53 (1.38, 1.71) | 0.55 |
| Mother's smoking | Never smoked<br>Smoked, not in pregnancy<br>Smoking in pregnancy | 1.00<br>0.80 (0.73, 0.88)<br>0.34 (0.30, 0.38) | 0.59 |
| Paternal smoking (ever smoked) | No<br>Yes | 1.00<br>0.63 (0.58, 0.69) | 0.56 |
| Behavioural difficulties score at 81 months | For each 1 point increase | 0.96 (0.95, 0.97) | 0.55 |
| Attainment at 11 years | For each 10% increase | 1.47 (1.43, 1.51) | 0.66 |
| Smoking at 14 years | No<br>Yes | 1.00<br>0.85 (0.70, 1.04) | 0.50 |
| Outcome: attainment score | For each 10% increase | 1.67 (1.61, 1.73) | 0.70 |

[1] Denominators in each analysis vary because vary because the variables come from different sources/questionnaires and have different completion rates.